\documentclass{ws-p8-50x6-00}
\usepackage{gribov}
\usepackage{amsmath}
\usepackage{psfig}

\def\lrang#1{\left\langle#1\right\rangle}
\def\abs#1{\left|#1\right|}
\def\acr{\alpha_{\mbox{\scriptsize crit}}}
\def\GeV{\mathop{\rm Ge\!V}}
\def\MeV{\mathop{\rm Me\!V}}
\def\cO#1{{\cal O}\left(#1\right)}
\def\Tr{\mathop{\rm Tr}}

\newcounter{hran}
\renewcommand{\thehran}{\arabic{hran}}

\def\bmini{\setcounter{hran}{\value{equation}}
\refstepcounter{hran}\setcounter{equation}{0}
\renewcommand{\theequation}{\thehran\alph{equation}}
\begin{eqnarray}}

\def\bminiG#1{\setcounter{hran}{\value{equation}}
\refstepcounter{hran}\setcounter{equation}{-1}
\renewcommand{\theequation}{\thehran\alph{equation}}
\refstepcounter{equation}\label{#1}\begin{eqnarray}}

\def\emini{\end{eqnarray}\relax\setcounter{equation}{\value{hran}}\renewcommand{\theequation}{\arabic{equation}}}

 \newskip\humongous \humongous=0pt plus 1000pt minus 1000pt
 \def\caja{\mathsurround=0pt} \def\eqalign#1{\,\vcenter{\openup1\jot
 \caja   \ialign{\strut \hfil$\displaystyle{##}$&$
 \displaystyle{{}##}$\hfil\crcr#1\crcr}}\,} \newif\ifdtup

\begin{document}

\title{Gribov program of understanding confinement}

\author{Yuri L.\ Dokshitzer}

\address{LPTHE, Universit\'e Paris--VI \&\ VII,}
\address{LPT, Universit\'e Paris--XI, }
\address{PNPI, Gatchina, 188350 St.\ Petersburg, Russia.}
\maketitle

\abstracts{}

\section{Introduction}

For V.N.~Gribov, the problem of confinement as the problem of
understanding the dynamics of vacuum fluctuations, and of the
structure of hadrons as physical states of the theory, was always
inseparable from the problem of understanding the physics of high
energy hadron scattering (the Pomeron picture).
Hand-written notes of one of his talks on confinement started with the
sentence
\begin{quote}
 {\em  Confinement --- older than quarks themselves.}
\end{quote}
and ended with 
\begin{quote}
... [the picture] {\em can be checked}\/ [in high energy scattering]
{\em on nuclei.}
\end{quote}
My aim is to introduce you to the Gribov scenario, and theory, of the
quark confinement. But first, a short excursion into the history of
high energy physics, and Gribov's place in it.\footnote{Based on the
preface to the book\cite{RBook}, written in cooperation with Leonid
Frankfurt.}

\section{High energy hadron scattering phenomena}

In the late fifties, when Gribov, then a young researcher at the Ioffe
Physico-Technical Institute, became interested in the physics of
strong hadron interactions, there was no consistent picture of of
high-energy scattering processes, not to mention {\em a theory}.
Apart from the Pomeranchuk theorem --- an asymptotic equality of
particle and antiparticle cross sections\cite{Pom} --- not much was
theoretically understood about processes at high energies.

\subsection{Asymptotic behaviour $s^{\alpha(t)}$}
Gribov's 1960 paper ``{\em Asymptotic behaviour of the scattering amplitude at
high energies}''
%
%
in which he proved an inconsistency of the so-called black-disk model
of diffractive hadron-hadron scattering, may be considered a first
building block of the modern theory of high-energy particle
interactions.\cite{1960}

Gribov's use of the so-called double-dispersion representation for the
scattering amplitude, suggested by S.\ Mandelstam back in
1958,\cite{Mand} demonstrated the combined power of the general
principles of relativistic quantum theory --- unitarity (conservation
of probability), analyticity (causality) and the relativistic nature
(crossing symmetry) --- as applied to high energy interactions.

By studying the analytic properties in the cross channels, he showed
that the imaginary part of the scattering amplitude in the form
\begin{equation} \label{blackdisk}
   A_1(s,t) = s\,f(t)
\end{equation}
that constituted the black-disk model of diffraction in the physical
region of $s$-channel scattering, contradicts the unitarity relation
for partial waves in the crossing $t$-channel. To solve the puzzle,
Gribov suggested the behaviour of the
amplitude (for large $s$ and finite $t$) in the general form
\begin{equation}
\label{Fst}
   A_1(s,t)= s^{q(t)}\,B_t(\ln s) \,,
\end{equation}
(equation (16) in \cite{1960}), where $B_t$ is a slow
(non-exponential) function of $\ln s$ (decreasing fast with $t$) and
$q(0)=1$ ensures the approximate constancy of the total cross section
with energy, $\sigma^{\mbox{\scriptsize tot}}(s)\simeq
\mbox{const}$.

In this first paper Gribov analyzed the constant exponent, $q(t)=1$,
and proved that the cross section in this case has to decrease at high
energies, $B_t(\ln s)< 1/\ln s$, to be consistent with the $t$-channel
unitarity.  He remarked on the possibility $q(t)\neq\mbox{const}$ as
``{\em extremely unlikely}'' since, considering the $t$-dependence of
the scattering amplitude, this would correspond to a strange picture
of the radius of a hadron infinitely increasing with energy.
He decided to ``{\em postpone the treatment of such rapidly changing
functions until a more detailed investigation is carried out}''.

He published the results of such an investigation the next year in the
letter to ZhETF ``{\em Possible asymptotic behaviour of elastic
scattering}''.
In his letter Gribov discussed the asymptotic behaviour ``{\em which
in spite of having a few unusual features is theoretically feasible
and does not contradict the experimental data}''.\cite{letter} Gribov
was already aware of the finding by T.~Regge\cite{Regge} that in
non-relativistic quantum mechanics
\begin{equation}\label{Reggeamp}
    A(s,t) \>\propto\> t^{\ell(s)}\,,
\end{equation}
in the unphysical region $|t|\gg s$ (corresponding to large imaginary
scattering angles $\cos\Theta\to\infty$), where $\ell(s)$ is the
position of the pole of the partial wave $f_\ell$ in the complex plane
of the orbital momentum $\ell$.

T.~Regge found that the poles of the amplitude in the complex
$\ell$-plane were intimately related with bound states/resonances.  It
is this aspect of the Regge behaviour that initially attracted the
most attention:
\begin{quote}
``{\em S.\ Mandelstam has suggested and emphasized repeatedly since
1960 that the Regge behaviour would permit a simple description of
dynamical states (private discussions). Similar remarks have been made
by R.~Blankenbecker and M.L.~Goldberger and by K.~Wilson}'' \hfill
(quoted from\cite{FGZ}).
\end{quote}
Gribov apparently learned about the Regge results from a paper by G.\
Chew and S.\ Frautschi of 1960\cite{CF60} which contained a {\em
footnote}\/ describing the main Regge findings.  In their paper, Chew
and Frautschi had advocated the standard black-disk diffraction model
(\ref{blackdisk}), and referred to Regge only in the context of the
``{\em connection between asymptotic behaviour in $t$ and the maximum
$\ell$ that interacts strongly}''.

The structure of the Regge amplitude (\ref{Reggeamp}) motivated Gribov
to return to the consideration of the case of the $t$-dependent
exponent in his general high-energy ansatz (\ref{Fst}) that was
dictated by $t$-channel unitarity.

By then M.~Froissart had already proved his famous theorem that limits
the asymptotic behaviour of the total cross sections,\cite{Froissart}
\begin{equation}\label{stot}
 \sigma^{\mbox{\scriptsize tot}} \propto s^{-1}\left| A_1(s,0)\right|
 \><\>  C\, \ln^2 s\,.
\end{equation}
Thus, having accepted $\ell(0)=1$ for the rightmost pole in the
$\ell$-plane as the condition ``{\em that the strongest possible
interaction is realized}'', Gribov formulated ``{\em the main
properties of such an asymptotic scattering behaviour}'':
\begin{itemize}
\item
  the total interaction cross section is constant at high energies,
\item
  the elastic cross section tends to zero as $1/\ln s$,
\item
  the scattering amplitude is essentially imaginary,
\item
the significant region of momentum transfer in elastic scattering
shrinks with energy increasing, $\sqrt{-t}\propto (\ln s)^{-1/2}$.
\end{itemize}
He also analysed the $s$-channel partial waves to show that for small
impact parameters $\rho<R$ their amplitudes fall as $1/\ln s$, while
the interaction radius $R$ increases with energy as
$\rho\propto\sqrt{\ln s}$.  He concluded:
\begin{quote}
``{\em this behaviour means that the particles become grey with
respect to high-energy interaction, but increase in size, so that the
total cross section remains constant}''.
\end{quote}
These were the key features of what has become known, quite
ironically, as the ``Regge theory'' of strong interactions at high
energies.
On the opposite side of the Iron Curtain, the basic properties of the
Regge pole picture of forward/backward scattering were formulated half
a year later by G.~Chew and S.~Frautschi in\cite{CF}.  In particular,
they suggested ``{\em the possibility that the recently discovered
$\rho$ meson is associated with a Regge pole whose internal quantum
numbers are those of an $I=1$ two-pion configuration}'', and
conjectured the universal high-energy behaviour of backward
$\pi^+\pi^0$, $K^+K^0$ and $pn$ scattering due to $\rho$-reggeon
exchange.  G.~Chew and S.~Frautschi also stressed that the
hypothetical Regge pole with $\alpha(0)=1$ responsible for forward
scattering possesses quantum numbers of the {\em vacuum}.

Dominance of the vacuum pole automatically satisfies the Pomeranchuk
theorem.  The name ``Pomeron'' for this vacuum pole was coined by
Murray Gell-Mann, who referred to Geoffrey Chew as an inventor.

{\em Shrinkage of the diffractive peak}\/ was predicted, and was
experimentally verified at particle accelerator experiments in Russia
(IHEP, Serpukhov), Switzerland (CERN) and the US (FNAL, Chicago), as
were the general relations between the cross sections of different
processes, that followed from the Gribov factorization
theorem.\cite{GP}

\subsection{Complex angular momenta in relativistic theory}

In non-relativistic quantum mechanics the interaction Hamiltonian
allows for scattering partial waves to be considered as analytic
functions of complex angular momentum $\ell$ (provided the interaction
potential is analytic in $r$).

Gribov's paper ``{\em Partial waves with complex orbital angular
momenta and the asymptotic behaviour of the scattering amplitude}''
%
%
showed that the partial waves with complex angular momenta can be
introduced in a relativistic theory as well,
on the basis of the Mandelstam double dispersion representation.

Here it is the unitarity in the crossing channel that {\em replaces}\/
the Hamiltonian and leads to analyticity of the partial waves in
$\ell$.  The corresponding construction is known as the
``Gribov-Froissart projection''.\cite{GF}

Few months later
%
%
Gribov demonstrated that the simplest (two-particle) $t$-channel
unitarity condition indeed generates the moving {\em pole}\/
singularities in the complex $\ell$-plane.  This was the {\em proof}\/
of the Regge hypothesis in relativistic theory.\cite{poles}

The ``Regge trajectories'' $\alpha(t)$ combine hadrons into families:
$s_h=\alpha(m_h^2)$, where $s_h$ and $m_h$ are the spin and the mass
of a hadron (hadronic resonance) with given quantum numbers (baryon
number, isotopic spin, strangeness, etc.).\cite{CF}  Moreover, at
negative values of $t$, that is in the physical region of the
$s$-channel, the very same function $\alpha(t)$ determines the
scattering amplitude, according to (\ref{Fst}).  It looks {\em as
if}\/ high-energy scattering was due to $t$-channel exchange of a
``particle'' with spin $\alpha(t)$ that varies with momentum transfer
$t$ --- the ``reggeon''.

Thus, the high-energy behaviour of the scattering process $a+b\to c+d$
is linked with the spectrum of excitations (resonances) of low-energy
scattering in the dual channel, $a+\bar{c}\to \bar{b}+d$.
This intriguing relation triggered many new ideas (bootstrap, the
concept of duality).
Backed by the mysterious {\em linearity}\/ of Regge trajectories
relating spins and squared masses of observed hadrons, the duality
ideas, via the famous Veneziano amplitude, gave rise to the concept of
hadronic strings and to development of string theories in general.

\subsection{Interacting Pomerons}

A number of theoretical efforts was devoted to understanding the
approximately constant behaviour of the total cross sections at high
energies.

To construct a full theory that would include the Pomeron trajectory
with the maximal ``intercept'' that respects the Froissart bound,
$\alpha_P(0)\!=\!1$, and would be consistent with unitarity and
analyticity proved to be very difficult.
This is because multi-Pomeron exchanges become essential, which
generate branch points in the complex plane of angular momentum
$\ell$. These branch point singularities were first discovered by
Mandelstam in his seminal paper of 1963\cite{Mcuts}.

%

Moreover, the study of particle production processes with large
rapidity gaps led V.N.\ Gribov, I.Ya.\ Pomeranchuk and K.A.\
Ter-Martirosyan to the concept of interacting reggeons.

By the end of the 60-s V.~Gribov had developed the general theory
known as Gribov Reggeon Calculus. He formulated the rules for
constructing the field theory of interacting Pomerons --- the Gribov
Reggeon Field Theory (RFT) --- and developed the corresponding diagram
technique.  Gribov RFT reduces the problem of high energy scattering
to a non-relativistic quantum field theory of interacting particles in
2+1 dimensions.

\subsection{Gribov partons and Feynman partons}

One of Gribov's most important contributions to high energy hadron
physics was the understanding of the space-time evolution of high
energy hadron-hadron and lepton-hadron processes, in particular the
nature of the reggeon exchange from the $s$-channel point of view.

In his lecture at the LNPI Winter School in 1973,\cite{lecture}
Gribov outlined the general phenomena and typical features that were
characteristic for high energy processes in any quantum field theory.
This lecture gives a perfect insight into Gribov's extraordinary way
of approaching complicated physical problems of general nature.  The
power of Gribov's approach lies in applying the universal picture of
fluctuating hadrons to both {\em soft}\/ and {\em hard}\/
interactions.

To understand the structure of hard (deep inelastic) photon--hadron
interactions Feynman suggested the idea of partons --- point-like
constituents of hadronic matter.

Gribov's partons are constituents of hadron matter, components of
long-living fluctuations of the hadron projectile, which are
responsible for soft hadron-hadron interactions: total cross sections,
diffraction, multiparticle production, etc.

Feynman defined partons in the infinite momentum frame to suppress
vacuum fluctuations whose presence would have undermined the notion of
the parton wave function of a hadron.\cite{Feynm}

Gribov's earlier work ``Interaction of $\gamma$-quanta and electrons
with nuclei at high energies''\cite{gammanucl} had been a precursor
to the famous Feynman paper.  Gribov described the photon interaction
in the rest frame of the target nucleus.  An incident real photon (or
a virtual photon in the deep inelastic scattering case) fluctuates
into hadrons before the target, at the longitudinal distance $L$
increasing with energy.  (For the $e^-p$ deep inelastic scattering
B.L.~Ioffe has shown that the assumption of Bjorken scaling implies
$L\sim 1/2xm_N$, with $x$ the Bjorken variable and $m_N$ the nucleon
mass.\cite{Ioffe})
  Therefore, at sufficiently large energy, when the fluctuation
  distance exceeds the size of the target, the photon no longer
  behaves as a point-like weekly interacting particle. Its interaction
  resembles that of a hadron and becomes ``black'', corresponding to
  complete absorption on a large nucleus.
This paper can be viewed as the generalization of the VDM (vector
dominance model) to high energy processes.

Being formally equivalent to Feynman's treatment, Gribov's approach is
better suited for the analysis of deep inelastic phenomena at very
small Bjorken $x$, where the interaction becomes actually strong, and
the perturbative QCD treatment is bound to fail.

Gribov diffusion in the impact parameter space giving rise to energy
increase of the interaction radius and to the reggeon exchange
amplitude, coexisting fluctuations as a source of branch cuts, duality
between hadrons and partons, a common basis for hard and soft elastic,
diffractive and inelastic process --- these are some of the key
features of high energy phenomena in quantum field theories, which are
still too hard a nut for QCD to crack.

\subsection{Gribov reggeon field theory}

Two best known applications of the Gribov RFT are
\begin{itemize}
\item
general quantitative relation between the shadowing phenomenon in
hadron-hadron scattering, the cross section of diffractive processes
and inelastic multi-particle production, known as
``Abramovsky-Gri\-bov-Kan\-chelli (AGK) cutting rules'',\cite{AGK} and
\item 
the essential revision by Gribov of the Glauber theory of nuclear
shadowing in hadron-nucleus interactions.\cite{GGlaub}
\end{itemize}

In 1968 V.N.\ Gribov and A.A.~Migdal demonstrated that the scaling
behaviour of the Green functions emerged in field theory in the strong
coupling regime.\cite{GM} Their technique helped to build the
quantitative theory of second order phase transitions and to analyse
critical indices characterizing the long-range fluctuations near the
critical point.

The problem of high energy behaviour of soft interactions remained
unsolved, although some viable options were suggested. In particular,
in ``{\em Properties of Pomeranchuk Poles, diffraction scattering and
asymptotic equality of total cross sections}''\cite{equal} Gribov has
shown that a possible consistent solution of the RFT in the so-called
{\em weak coupling}\/ regime calls for the formal asymptotic equality
of {\em all}\/ total cross sections of strongly interacting particles.

Gribov's last work in this subject was devoted to the intermediate
energy range and dealt with interacting hadron fluctuations.\cite{heavyP}

The study of the {\em strong coupling}\/ regime of interacting
reggeons (pioneered by A.B.\ Kai\-da\-lov and K.A.Ter-Martirosyan) led
to the introduction of the {\em bare}\/ Pomeron with
$\alpha_P(0)\!>\!1$. The RFT based on $t$-channel unitarity should
enforce the $s$-channel unitarity as well. The combination of
increasing interaction radius and the amplitudes in the impact
parameter space which did not fall as $1/\ln s$ (as in the one-Pomeron
picture) led to logarithmically increasing asymptotic cross sections,
resembling the Froissart regime (and respecting the Froissart bound
(\ref{stot})).
The popularity of the notion of the bare Pomeron with
$\alpha_P(0)\!>\!1$ is based on experiment (increasing total hadron
cross sections).  Psychologically, it is also supported by the
perturbative QCD finding that the (small) scattering cross section of
two small-transverse-size objects should increase with energy in a
power-like fashion in the restricted energy range (the so-called hard
BFKL regime\cite{BFKL}).

\subsection{Reggeization and the Pomeron singularity in gauge theories}

In the mid-sixties V.N.\ Gribov initiated the study of double
logarithmic asymptotics of various processes in quantum
electrodynamics (QED), making use of the powerful technique he had
developed for the analysis of the asymptotic behaviour of Feynman
diagrams in the limit $s\to \infty$ and
$t=\mbox{const}$.\cite{Gribov_lectures}

In particular, in 1975 V.N.\ Gribov, L.N.\ Lipatov and G.V.\ Frolov
studied the high energy behaviour of QED processes from the point of
view of ``Regge theory''. High energy scattering amplitudes with
exchange of an {\em electron}\/ in the $t$-channel acquire, in higher
orders in QED coupling, a characteristic behaviour $A\propto s^{j(t)}$
with $j(m_e^2)=1/2$, that is, electron becomes a part of the Regge
trajectory: {\em reggeizes}. (So do quarks and gluons in QCD; V.S.\
Fadin, L.L.\ Frankfurt, L.N.Lipatov, V.E.\ Sherman, 1976.)  For the
{\em vacuum channel}, however, Gribov, Lipatov and Frolov
found\cite{qedjplane} that the rightmost singularity in the complex
$j$-plane is not a moving pole (as it is, e.g., for electron) but,
instead, a fixed branch point singularity positioned to the right from
unity, $j=1+c\alpha^2>1$.
This was a precursor of a similar result found later by Fadin, Lipatov
and Kuraev\cite{BFKL} in non-Abelian theories, and QCD in particular.
The problem of apparent anti-Froissart behaviour of the perturbative
``hard Pomeron'' in QCD still awaits resolution.

With the advent of quantum chromodynamics as a microscopic theory of
hadrons and their interactions, the focus of theoretical studies has
temporarily shifted away from Gribov-Regge problematics to ``hard''
small-distance phenomena.

\section{Gribov as QCD apprentice}

V.N.~Gribov became interested in non-Abelian field theories relatively
late, in 1976. His very first study, as a QCD apprentice, produced
amazing results.

In February 1977 in the proceedings of the 12$^{th}$ PNPI Winter
School he published two lectures which were to change forever the
non-Abelian landscape.

In the first lecture ``{\em Instability of non-Abelian gauge fields
and impossibility of the choice of the Coulomb gauge}''\cite{BH1} he
showed that the three-dimensional transversality condition
\begin{equation}
\label{eq:Btr}
 \left(\bpartial \cdot {\bf B}\right) 
  \equiv   \frac{\partial {B}_i}{\partial x_i } =
  0\,, \qquad i=1,2,3
\end{equation}
which is usually imposed on the fields to describe massless vector
particles (the Coulomb gauge), {\em does not}\/ solve the problem of
gauge fixing. Due to essential non-linearity of gauge transformation,
a ``transverse'' field \eqref{eq:Btr} may actually happen to be a pure
gauge field which should not be counted as a physical degree of
freedom.

Gribov explicitly constructed such ``transversal gauge fields'' for
the $SU(2)$ gauge group and showed that the {\em uncertainty}\/ in
gauge fixing arises when the effective magnitude of the field becomes
large,
\[
    L\cdot B \>\sim \frac{1}{g_s},
\]
or, in other words, when the effective interaction strength (QCD
coupling) becomes of the order of unity, that is, in the
non-perturbative region.

He also gave an elegant physically transparent explanation of the {\em
anti-screening}\/ phenomenon within the Hamiltonian approach, which had
been observed by I.~Khriplovich back in 1969\cite{Julik} in the
ghost-free Coulomb gauge and then (re)discovered and coined as the
{\em asymptotic freedom}\/ by D.\ Gross \& F.\ Wilczek and H.D.\
Politzer in 1973.

In the Hamiltonian language, there are (or, better to say, seem to be)
$N^2-1$ massless ${\bf B}$ fields (transverse gluons), and, as in QED,
an additional Coulomb field mediating interaction between colour
charges. Unlike QED, the non-Abelian Coulomb quantum has a colour
charge of its own. Therefore, traversing the space between two
external charge, it may virtually decay into two transverse fields,
\[
   {\bf 0} \>\to\> \bperp + \bperp \>\to\> {\bf 0}\,, 
\]   
or into a $q\bar{q}$ pair 
\[ 
   {\bf 0} \>\to\> {\bf q} + \bar{\bf q} \>\to\> {\bf 0}\,, 
\]
in the same manner as a QED Coulomb quantum fluctuates in the vacuum
into an $e^+e^-$ pair.  Both these effects lead to {\em screening}\/
of the colour charge of the external sources, in a perfect accord with
one's physical intuition. The fact that there are ``physical'' fields
in the intermediate state --- quarks and/or transverse gluons ---
fixes the sign of the virtual correction to correspond to {\em
screening}, via the unitarity relation.

Technically speaking, these virtual decay processes contribute to the
QCD $\beta$-function as
\bminiG{beta}
\label{eq:betaphys}
\left\{ \frac{d\,\alpha_s^{-1}(R)}{d\ln R}\right\}^{\mbox{\scriptsize phys}}  
\>\propto\> \frac13\, N \,+\,  \frac{2}{3}\,n_f\,,
\end{eqnarray}
that is, make the effective (running) coupling {\em decrease}\/ at
{\em large}\/ distances $R$ between the external charges.

Where then the {\em anti-screening}\/ comes from?  It originates from
another, specifically non-Abelian, effect namely, interaction of the
Coulomb quantum with the field of ``zero-fluctuations'' of transverse
gluons in the vacuum,
\[
  \Big[ {\bf 0}\>+ \bperp \>\to\> {\bf 0}\Big]^n. 
\]
In a course of such multiple rescattering, the Coulomb quantum preserves
its identity as an {\em instantaneous}\/ interaction mediator, and
therefore is not affected by the unitarity constraints. 
Statistical average over the transverse vacuum fields in the second
order of perturbation theory ($n=2$; the $n=1$ contribution vanishes
upon averaging) results in an additional contribution to the
Coulomb interaction energy which, translated into the running coupling
language, gives
\begin{eqnarray}
\label{eq:betastat}
\left\{ \frac{d\,\alpha_s^{-1}(R)}{d\ln R}\right\}^{\mbox{\scriptsize stat}}  
\>\propto\> -4\, N 
\emini
Taken together, the two contributions \eqref{beta} combine into the
standard QCD $\beta$-function.
An anti-intuitive minus sign in \eqref{eq:betastat} has its own simple
explanation. It is of the same origin as the minus sign in the shift
of the energy of the ground state of a quantum-mechanical system under
the second order of perturbation:
\[
 \delta E \equiv E-E_0 
 = \sum_n \frac{\abs{\lrang{0|\delta V|n}}^2}{E_0-E_n} \> < \> 0\,.
\] 
The r\^ole of perturbation $\delta V$ is played here by transverse
vacuum fields. 

Propagation of a Coulomb field $C^a$ in the ``external'' transverse gluon
field ${\bf B}^a_\perp$
is described by the operator which resembles the Faddeev--Popov ghost
propagator,
\begin{equation}
\label{eq:FPgh}
  G^{-1}[{\bf B}_\perp]  = ({\bf D}\cdot\bpartial)   \equiv 
\partial_i^2 
  + ig_s\left[B_{i\perp},\partial_i  \right] ,
\end{equation}
with ${\bf D}$ the covariant derivative,
\[
 D_i[{\bf B}_\perp]\, \psi \>=\> \partial_i\,\psi  
    + ig_s\left[B_{i\perp},\psi\right].
\]
For small vacuum fields, $g_s{\bf B}_\perp/\bpartial\ll 1$, expanding
perturbatively the Coulomb propagator $G$ to the second order in $g_s$
produces the perturbative anti-screening effect as stated
in~\eqref{eq:betastat}.

If we take, however, the gluon field in the QCD vacuum as large as
\[
  g_s{\bf B}_\perp/\bpartial \sim g_s{\bf B}_\perp\cdot L \sim 1\,,
\] 
a qualitatively new phenomenon takes place namely, the Coulomb (ghost)
propagator may become singular:
\begin{equation}
\label{eq:zero}
  G^{-1}[{\bf B}_\perp]C_0 \>=\>  C_0 \left( \partial_i^2 C_0 
  + ig_s\left[B_{i\perp},\partial_i C_0 \right] \right) \>=\>0\,. 
\end{equation}
Such a ``zero-mode'' solution $C_0$ in the external field is a sign of
an infrared instability of the theory. On the other hand, the
``surface'' $G^{-1}[{\bf B}_\perp]=0$ in the functional ${\bf
B}_\perp$--space is that very border beyond which the solution of the
gauge-fixing equation \eqref{eq:Btr} becomes copious.

The fact that the Coulomb propagator develops singularity does not
necessarily mean that the Faddev--Popov ghost ``rises from the dead''
by pretending to propagate as a particle. It rather tells us that we
have failed to formulate the quantum theory of interacting vector
fields, to properly fix physical degrees of freedom.

The existence of ``Gribov copies'' means that the standard Faddeev--Popov
prescription for quantizing non-Abelian gauge theories is, strictly
speaking, incomplete and should be modified.

Gribov addressed the problem of possible modification of the
quantization procedure in the second lecture ``Quantization of
non-Abelian gauge theories''\cite{BH2}.
The paper under the same title based on these two lectures is now a
universally accepted (though disturbing) truth and during 25 odd years
since its appearance in 1978\cite{Quant} was being cited more than 20
times per year, in average.
It goes without saying, that this paper was initially rejected by a
NPB referee, who wrote (as far as I can remember) that {\em the author raises
the problem of confinement which problem had been already solved and
isn't worth talking about}.

To properly formulate non-Abelian field dynamics, Gribov suggested
to limit the integration over the fields in the functional integral to
the so-called {\em fundamental domain}, where the Faddeev--Popov
determinant is strictly positive (the region in the functional space
of transverse fields ${\bf B}_\perp$ {\em before}\/ the first zero
mode \eqref{eq:zero} emerges).

In the second lecture, Gribov produced qualitative arguments in favour
of the characteristic modification of the gluon propagator, due to the
new restriction imposed on the functional integral. Effective
suppression of large gluon field results, semi-quantitatively, in
an infrared singular polarization operator
\bminiG{Gprop}
\label{eq:Gprop1}
 D^{-1}(k) \simeq k^2 + \frac{\lrang{G^2}}{k^2},
\end{eqnarray}
which coincides with the perturbative gluon Green function at large
momenta (small distances), $D(k)\propto k^{-2}$ but makes it {\em
vanish}\/ of at $k=0$, instead if having a pole corresponding to
massless gluons:
\begin{eqnarray}
\label{eq:Gprop2}
  D(k) \propto \frac{k^2}{k^4 + \lrang{G^2}} . 
\emini
The new non-perturbative parameter $\lrang{G^2}$ in \eqref{Gprop} has
dimension (and the meaning) of the familiar vacuum gluon condensate.

Literally speaking, the ansatz \eqref{eq:Gprop2} cannot be correct
since such a Green function would violate causality.\footnote{It is
unfortunate, therefore, that the form
\eqref{eq:Gprop2} which Gribov suggested and discussed for
illustrative purposes only, is often referred to in the literature as
``the Gribov propagator''.}
In reality, the gluon (as well as the quark) propagator should have a
more sophisticated analytic structure with singularities on unphysical
sheets, which would correspond, in the standard field-theoretical
language, to {\em unstable}\/ particles.

In spite of many attempts, the problem of {\em Gribov copies}\/
(``Gribov horizon'', ``Gribov uncertainties'') remains open today and
plagues the dynamics of any essentially non-linear gauge invariant
systems.  A promising recent attempt to implement the Gribov
fundamental domain restriction in pure gluodynamics, and its
prehistory, can be found in\cite{BZ}.

Gribov did not pursue this goal himself not because of severe
difficulties in describing the fundamental domain in the functional
space: he always had his ways around technical obstacles.
He convinced himself (though not yet the physics community at large)
that the solution to the confinement problem lies not in the
understanding of the interaction of ``large gluon fields'' but instead
in the understanding of how the QCD dynamics can be arranged as to prevent
the non-Abelian fields from growing real big.

\section{Gribov light quark confinement}

By 1990 V.~Gribov formulated the ``light-quark confinement scenario''
which was essentially based on the existence of two very light
(practically) massless quarks in the theory.

\subsection{Super-critical binding of fermions}

As a result of the search for a possible solution of the confinement
puzzle, which he pursued with unmatched intensity and depth for more
than 10 years after 1977, Gribov formulated for himself the key
ingredients of the problem and, correspondingly, the lines to approach
it:
\begin{itemize}
\item
    The question of interest is not of ``a'' confinement, but that of
    ``the'' confinement in the real world, namely, in the world with
    two very light quarks ($u$ and $d$) whose Compton wave lengths are
    much larger than the characteristic confinement scale ($m_q\sim
    5-10\,\MeV \ll 1\,\GeV$).
\item
    No mechanism for binding massless {\em bosons}\/ (gluons) seems to
    exist in QFT, while the Pauli exclusion principle may provide
    means for binding together massless {\em fermions}\/ (light
    quarks).
\item
    The problem of ultraviolet regularization may be more than a
    technical trick in a QFT with apparently infrared-unstable
    dynamics: the ultraviolet and infrared regimes of the theory may
    be closely linked. Example: the pion field as a Goldsone boson
    emerging due to spontaneous chiral symmetry breaking (short
    distances) and as a quark bound state (large distances).
\item
    The Feynman diagram technique has to be reconsidered in QCD if one
    goes beyond trivial perturbative correction effects.  
    Feynman's famous $i\epsilon$ prescription was designed for (and is
    applicable only to) the theories with stable perturbative vacua.
    To understand and describe a physical process in a confining
    theory, it is necessary to take into consideration the response of
    the vacuum, which leads to essential modifications of the quark
    and gluon Green functions. 
\end{itemize}

There was a deep reason for this turn, which Gribov formulated in the
following words:
\begin{quote} 
  ``I found I don't know how to bind massless {\em bosons}''
\end{quote}
(read: how to dynamically construct {\em glueballs}\/).  As for
fermions, there is a corresponding mechanism provided by the
Fermi--Dirac statistics and the concept of the ``Dirac sea''.
Spin--1/2 particles, even massless which are difficult to localize,
can be held together simply by the fact that, if pulled apart, they
would correspond to the free-fermion states that are {\em occupied}\/
as belonging to the Dirac sea. In a pure perturbative
(non-interacting) picture, the empty fermion states have {\em positive
energies}, while the {\em negative-energy}\/ states are all filled.
With account of interaction the situation may change, {\em provided}\/
two {\em positive-energy}\/ fermions (quarks) were tempted to form a
bound state with a {\em negative}\/ total energy. 
In such a case, the true vacuum of the theory would contain {\em
positive kinetic energy}\/ quarks hidden inside the {\em negative
energy}\/ pairs, thus preventing positive-energy quarks from flying
free.

A similar physical phenomenon is known in QED under the name of
super-critical binding in ultra-heavy nuclei.
Dirac energy levels of an electron in an external static field created
by the large electric charge $Z>137$ become complex. This means
instability: classically, the electron ``falls on the centre''.  In
QFT the instability develops when the energy $\epsilon$ of an empty
atomic electron level falls, with increase of $Z$, below $-m_ec^2$.
An $e^+e^-$ pair pops up from the vacuum, with the vacuum electron
occupying the level: the super-critically charged ion decays into an
``atom'' (the ion with the smaller charge, $Z-1$) and a real positron
$$
 A_Z \>\Longrightarrow A_{Z-1} + e^+\,, \qquad \mbox{for}\>
 Z>Z_{\mbox{\scriptsize crit.}}
$$
In the QCD context, the increase of the running quark-gluon coupling
at large distances replaces large $Z$ of the QED problem.

Gribov generalized the problem of super-critical binding in the field
of an infinitely heavy source to the case of two massless fermions
interacting via Coulomb-like exchange.  He showed that in this case
the super-critical phenomenon develops much earlier.

In the Lund University preprint\cite{Lund} Gribov has shown that a
{\em pair}\/ of light fermions interacting in a Coulomb-like manner
develops super-critical behaviour if the 
coupling hits a definite critical value
\begin{equation}
\label{eq:acr}
 \frac{\alpha}{\pi} > \frac{\acr}{\pi} = 1-\sqrt{\frac{2}{3}}\,.
\end{equation}
In QCD, with account of the colour Casimir operator, the criticality
condition translates into
\begin{equation}
\label{eq:acrQCD}
  \frac{\acr}{\pi} = C_F^{-1}\left[\, 1-\sqrt{\frac{2}{3}}\,\right]
  \simeq 0.137\,.
\end{equation}
This number, apart from being easy to remember, has another
interesting property: it is numerically {\em small}, which opens  
%
an intriguing, and perfectly heretic, possibility of understanding,
and describing (at least semi-quantitatively) the full wealth of
hadron spectra and interactions in the perturbative language of quarks
and gluons.

\subsection{Gribov equation}

Gribov constructed the equation for the quark Green function as an
approximation to the general corresponding Schwinger-Dyson
equation.

This approximation took into account the most singular
(logarithmically enhanced) infrared and ultraviolet effects due to
quark-gluon interactions and resulted in a non-linear differential
equation which possesses a rich non-perturbative structure.

An amazing simplicity of the Gribov construction makes one wonder,
why such an equation had not been discovered 15--20 years earlier when
a lot of effort was applied in a search for non-perturbative phenomena
of the super-conductivity type in QED (Nambu--Jona-Lasinio;
Baker--Johnson; Fomin--Miransky et al.; \ldots).

Take the first order self-energy diagram (which is easier for you to
imagine than for me to draw): a fermion (quark/electron) with momentum
$q$ virtually decays into a quark (electron) with momentum $q'$ and a
massless vector boson (gluon/photon) with momentum $q-q'$:
\begin{equation}
\label{eq:Sigma}
  \Sigma(q) = [C_F]\frac{\alpha}{\pi}
  \int \frac{d^4q'}{4\pi^2 i} \left[\, \gamma_\mu \,
  G(q')\, \gamma_\mu\, \right] \> D(q-q') ,\qquad D_0(k) =
  \frac{1}{k^2+i\epsilon},
\end{equation}
with $G$ and $D$ the fermion and boson propagators, respectively.
This Feynman diagram diverges linearly at $q'\to\infty$. To kill the
ultraviolet divergences (both linear and logarithmic), it suffices to
{\em differentiate}\/ it twice over the external momentum $q$.

The first Gribov's observation was that $1/k^2$ happens to be
the Green function of the four-dimensional Laplace operator, 
\[
   \partial^2 \frac{1}{(q-q')^2+i\epsilon} = -4\pi^2\, i\delta(q-q'),
   \qquad \partial_\mu \equiv \frac{\partial }{\partial q_\mu},
\]
where $\partial_\mu$ now denotes the momentum
differentiation. Therefore, the operation $\partial^2$ applied to
\eqref{eq:Sigma} takes away the internal momentum integration and
leads to an expression which is {\em local}\/ in the momentum space, 
\begin{equation}
\label{eq:Born}
   \partial^2 \Sigma_1(q) \>=\> [C_F]\frac{\alpha}{\pi}\, \gamma_\mu \,
    G_0(q)\, \gamma_\mu\,, \qquad k=q-q'=0\,.
\end{equation}
This is the ``Born approximation''. With account of higher order
radiative corrections, the first thing that happens is that the bare
fermion propagator $G_0$ dresses up, $G_0(q)\to G(q)$, and so do the
Born vertices $\gamma_\mu\to \Gamma_\mu(q,q,0)$. The second crucial
observation was that the exact vertex function $\Gamma_\mu(q,q-k,k)$
describing the emission of a {\em zero momentum}\/ vector boson,
$k_\mu\equiv 0$, is not an independent entity but is related with the
fermion propagator by the Ward identity,
\begin{equation}
   \Gamma_\mu(q,q,0) = -\partial_\mu G^{-1}(q)\,,
\end{equation}
which statement is {\em literally}\/ true in Abelian theory (QED),
and, after some reflection, {\em can be made}\/ true in the non-Abelian
case (QCD) as well.\footnote{I successfully fought the temptation to
call it an ``illiterally true'' statement.}

Thus, we have arrived to the Gribov equation for the quark Green
function\cite{Lund,conf1}
\begin{equation}
\label{eq:BHeq}
  \partial^2 G^{-1}(q) = g\, \partial_\mu G^{-1}(q)\, G(q)\,
  \partial_\mu G^{-1}(q) +
\ldots , \qquad  g\>\equiv\> C_F\frac{\alpha_s}{\pi},
\end{equation}
where $\ldots$ stand for less singular $\cO{g^2}$ integral terms.

Yet another set of higher order corrections makes the coupling run,
$g\to g(q^2)$. In the $\abs{q^2}\to\infty$ limit the QCD coupling
vanishes due to the asymptotic freedom, and
\eqref{eq:BHeq} turns into the free equation, $\partial^2 G^{-1}=0$,
whose solution has the form
\begin{equation}
 G^{-1}(q) = Z^{-1} \left[\, (m-\hat{q}) + \frac{\nu_1^3}{q^2}
 +\frac{\nu_2^4\,\hat{q}}{q^4} \,\right].
\end{equation}
This general perturbative solution has two new arbitrary parameters
$\nu_1$ and $\nu_2$ in addition to the familiar two (mass $m$ and the
wave function renormalization constant $Z$), since the master equation
is now the {\em second}\/ order differential equation, unlike the
standard renormalization group (RG) approach.

Since the new terms are singular at small $q$, in QED, for example, we
simply drop them as unwanted, thus returning to the RG structure. Such
a prescription, however, exploits the knowledge that nothing dramatic
happens in the infrared domain, so that the real electron in the
physical spectrum of the theory, whose propagation we seek to
describe, is inherently that very same object that we put into the
Lagrangian as a fundamental bare field.

In an infrared unstable theory (QCD) we better wait and see.

Indeed, Gribov found that if the coupling in the infrared region
exceeded the critical value \eqref{eq:acrQCD}, a bifurcation occurred in
\eqref{eq:BHeq} and a new phase emerged corresponding to {\em spontaneous
breaking of the chiral symmetry}.

\subsection{Quarks, pions and confinement}
As far as {\em confinement}\/ is concerned, the approximation
\eqref{eq:BHeq} turned out to be insufficient. A numerical 
study of the Gribov equation carried out by Carlo Ewerz
showed\cite{Carlo}
that the corresponding quark Green function does not possess an
analytic structure that would correspond to a confined object.

Given the dynamical chiral symmetry breaking, however, the Goldstone
phenomenon takes place bringing pions to life. 
In his last paper\cite{conf2}
Gribov argued that the effects that Goldstone pions induce, in turn,
on the propagation of quarks is likely to lead to confinement of light
quarks and, as a result, to confinement of any colour states.

The approximate equation for the Green function of a massless
quark, which accommodates a feed-back from Goldstone pions
reads\cite{conf2}
\begin{equation}
\label{eq:BHmod}
\eqalign{
 \partial^2 G^{-1}(q) =&\> g(q)\, \partial_\mu G^{-1}(q)\, G(q)\,
  \partial_\mu G^{-1}(q) \cr
  &- \frac{3}{16\pi^2 f_\pi^2}
\left\{i\gamma_5,G^{-1}(q)\right\}\,G(q)\, \left\{ i\gamma_5,G^{-1}(q)\right\}.
}\end{equation} 
The modified Gribov equation \eqref{eq:BHmod} still awaits a detailed
study aiming at the analytic structure of its solutions.

Another important open problem is to construct and to analyse a
similar equation in the gluon sector, from which a consistent picture
of the coupling $g(q)$ rising above the critical value in the infrared
momentum region should emerge.

It is important to notice that since pions have emerged {\em
dynamically}\/ in the theory, their coupling to quarks is not
arbitrary but is tightly linked with the quark propagator itself
(search for an anti-commutator of $\gamma_5$ with $G^{-1}$ in
\eqref{eq:BHmod}).  Moreover, the pion--axial current transition
constant $f_\pi$ is not arbitrary either, but has to satisfy a
definite relation which, once again, is driven by the behaviour of the
exact quark Green function:
\begin{equation}
\label{eq:fpi}
\eqalign{
f_\pi^2 =&\> \frac18 \int\frac{d^4q}{(2\pi)^4i} \Tr\left[\>
\left\{i\gamma_5,G^{-1}\right\}\,G\, \left\{ i\gamma_5,G^{-1}\right\} G
\left(\partial_\mu G^{-1}\,G\right)^2\>\right] \cr
&+ \frac1{64\pi^2f_\pi^2} \int \frac{d^4q}{(2\pi)^4i} \Tr\left[\> 
\left(\left\{i\gamma_5,G^{-1}\right\}\,G\right)^4\>\right].
}\end{equation}

The second of the two papers\cite{conf1,conf2} concluding Gribov's
study of the light-quark super-critical confinement theory remained
unfinished.  It ends abruptly in the middle of the discussion of the
most intriguing question, namely, what is the meaning, and practical
realization, of unitarity in a confining theory.

Gribov works on gauge theories and, in particular, all his papers,
talks and lectures devoted to anomalies and the QCD confinement
(including some lectures that were translated from Russian for the
first time)
were collected in the book\cite{RBook} due to be published in Moscow
by the end of 2001.

From these papers, an interested reader will be able to derive the
Gribov equation and to study the properties of its (perturbative and
non-perturbative) solutions, as well as to formulate the open problems
awaiting analysis and resolution (markedly, how to construct an
equation for the running coupling, similar to that for the quark Green
function).  Gribov lectures will also give an opportunity to grasp the
physical picture of the super-critical QCD binding which includes the
notion of an ``inversely populated'' Dirac sea of light quarks, and to
think about phenomenological aspects, and verification, of the Gribov
confinement scenario.

Speaking of Gribov heritage, another must-to-have for each
university/educational body in theoretical physics is Gribov lectures
on Quantum Electrodynamics which can be considered, and used as, an
introductory course into relativistic field theory in general.\cite{QED}

\begin{figure}
 \psfig{file=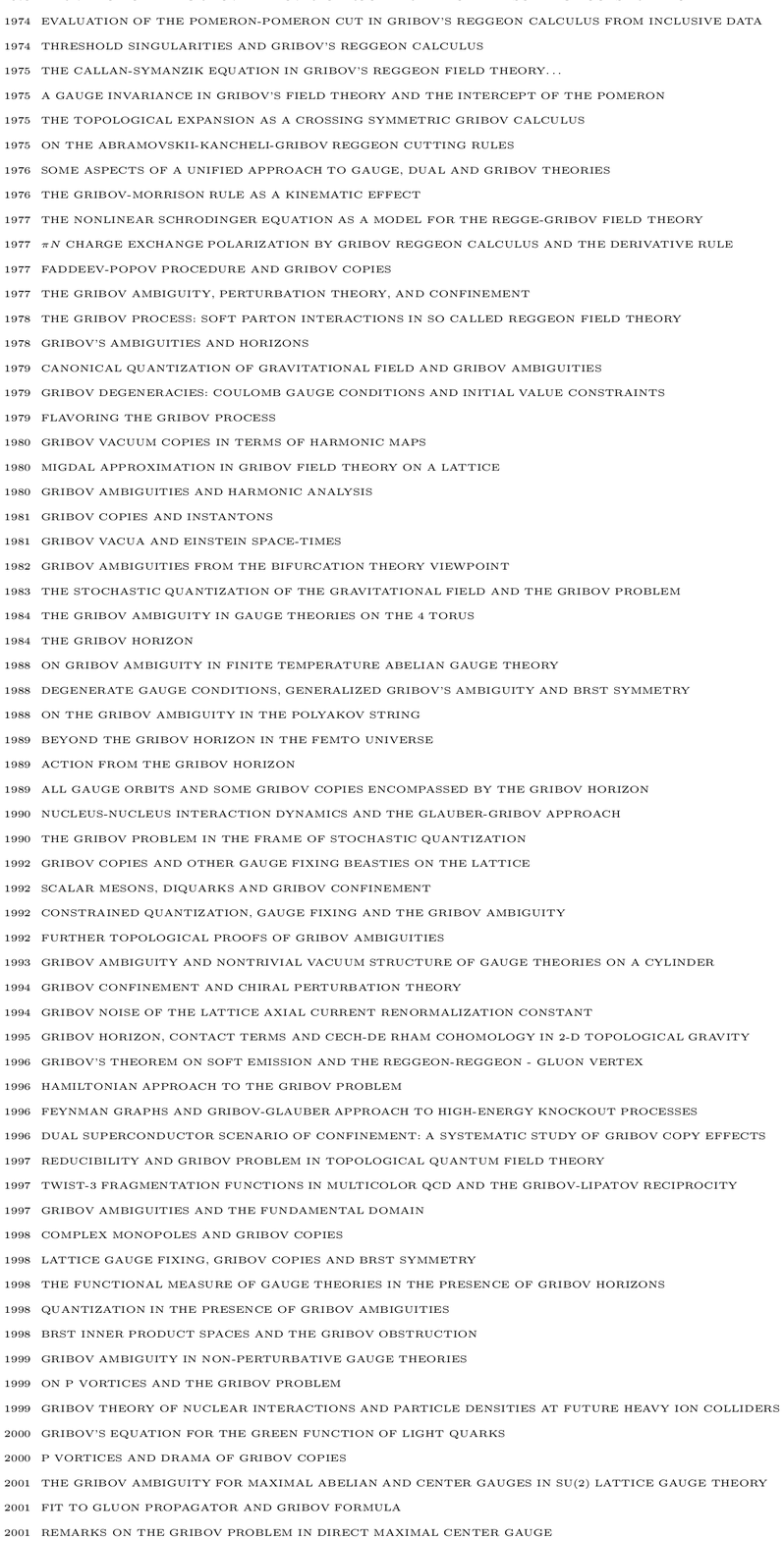,height=18cm}
\end{figure}

\end{document}